\documentclass{llncs}
\usepackage{amsmath,amsfonts,amssymb,algorithm2e,rotating}
\usepackage{graphicx}

\def \ie {i.e.~}
\def \NP {$\mathcal{NP}$}

\begin{document}
\title{A tabu search heuristic for the Equitable Coloring Problem}
\author{I. M\'endez D\'iaz\inst{3}, G. Nasini\inst{1,2} \and D. Sever\'in\inst{1,2}}
\institute{Facultad de Ciencias Exactas, Ingenier\'ia y Agrimensura \\ Universidad Nacional de Rosario, Argentina \\
\email{\{nasini,daniel\}@fceia.unr.edu.ar} \and CONICET, Argentina
\and Facultad de Ciencias Exactas y Naturales \\ Universidad de Buenos Aires, Argentina \\ \email{imendez@dc.uba.ar}}

\maketitle

\begin{abstract}
%The Equitable Coloring Problem is a variant of the Graph Coloring Problem, where the sizes of two arbitrary color classes
%differ in at most one unit.
%This paper describes a new tabu search heuristic for the Equitable Coloring Problem. This algorithm is an adaptation of the
%dynamic \textsc{TabuCol} version of Galinier and Hao.
%Computational experiments are carried out in order to find the best combination of parameters involved in the tabu search
%heuristic, and to show its good performance over benchmark instances.
The \emph{Equitable Coloring Problem} is a variant of the Graph Coloring Problem where the sizes of two arbitrary color classes differ in at most one unit. This additional condition, called equity constraints, arises naturally in several applications.
Due to the hardness of the problem, current exact algorithms can not solve large-sized instances. Such instances must be addressed only
via heuristic methods.

In this paper we present a tabu search heuristic for the Equitable Coloring Problem. This algorithm is an adaptation of the
dynamic \textsc{TabuCol} version of Galinier and Hao. In order to satisfy equity constraints, new local search criteria are given. 

Computational experiments are carried out in order to find the best combination of parameters involved in the dynamic tenure of the
heuristic.

Finally, we show the good performance of our heuristic over known benchmark instances.
\end{abstract}

\begin{keywords}
equitable coloring $\cdot$ tabu search $\cdot$ combinatorial optimization
\end{keywords}

\section{Introduction} \label{SINTRO}

The \emph{Graph Coloring Problem} (GCP) is a very well-studied \NP-Hard problem since it models many applications such as
scheduling, timetabling, electronic bandwidth allocation and sequencing problems.

Given a simple graph $G = (V, E)$, where $V$ is the set of vertices and $E$ is the set of edges, a \emph{$k$-coloring of $G$}
is a partition of $V$ into $k$ sets $V_1, V_2, \ldots, V_k$, called \emph{color classes}, such that the endpoints of any edge
lie in different color classes. The GCP consists of finding the minimum number $k$ such that $G$ admits a $k$-coloring,
called the \emph{chromatic number} of $G$ and denoted by $\chi(G)$.

Some applications impose additional restrictions. For instance, in scheduling problems, it may be required to ensure the
uniformity of the distribution of workload employees. Suppose that a set of tasks must be assigned to a set of workers so that
pairs of tasks may conflict each other, meaning that they should not be assigned to the same worker. The problem is modeled by
building a graph containing a vertex for every task and an edge for every conflicting pair of tasks. Workers are represented by colors.
Then, in order for a coloring of this graph to represent a valid assignment of tasks to workers, the same number of tasks must be assigned to each worker.
Since this is impossible when the number of tasks is not divisible by the number of workers, one can ask for the number of tasks
assigned to two arbitrary workers can not differ by more than one.
It is called \emph{equity constraint} and the resulting problem is called \emph{Equitable Coloring Problem} (ECP).

ECP was introduced in \cite{MEYER}, motivated by an application concerning \emph{garbage collection} \cite{EXAMPLE2}. 
Other applications of the ECP concern \emph{load balancing problems} in multiprocessor machines \cite{EXAMPLE3}
and results in \emph{probability theory} \cite{EXAMPLE1}. An introduction to ECP and some basic results are provided in
\cite{KUBALE}.

Formally, an \emph{equitable $k$-coloring} (or just $k$-eqcol) of a graph $G$ is a $k$-coloring 
satisfying the \emph{equity constraint}, \ie the size of two color classes can not differ by more than one unit.
The \emph{equitable chromatic number} of $G$, $\chi_{eq}(G)$, is the
minimum $k$ for which $G$ admits a $k$-eqcol. The ECP consists of finding $\chi_{eq}(G)$ which is an \NP-Hard problem \cite{KUBALE}.

\begin{figure}[h]
\begin{center}
\includegraphics[scale=0.2]{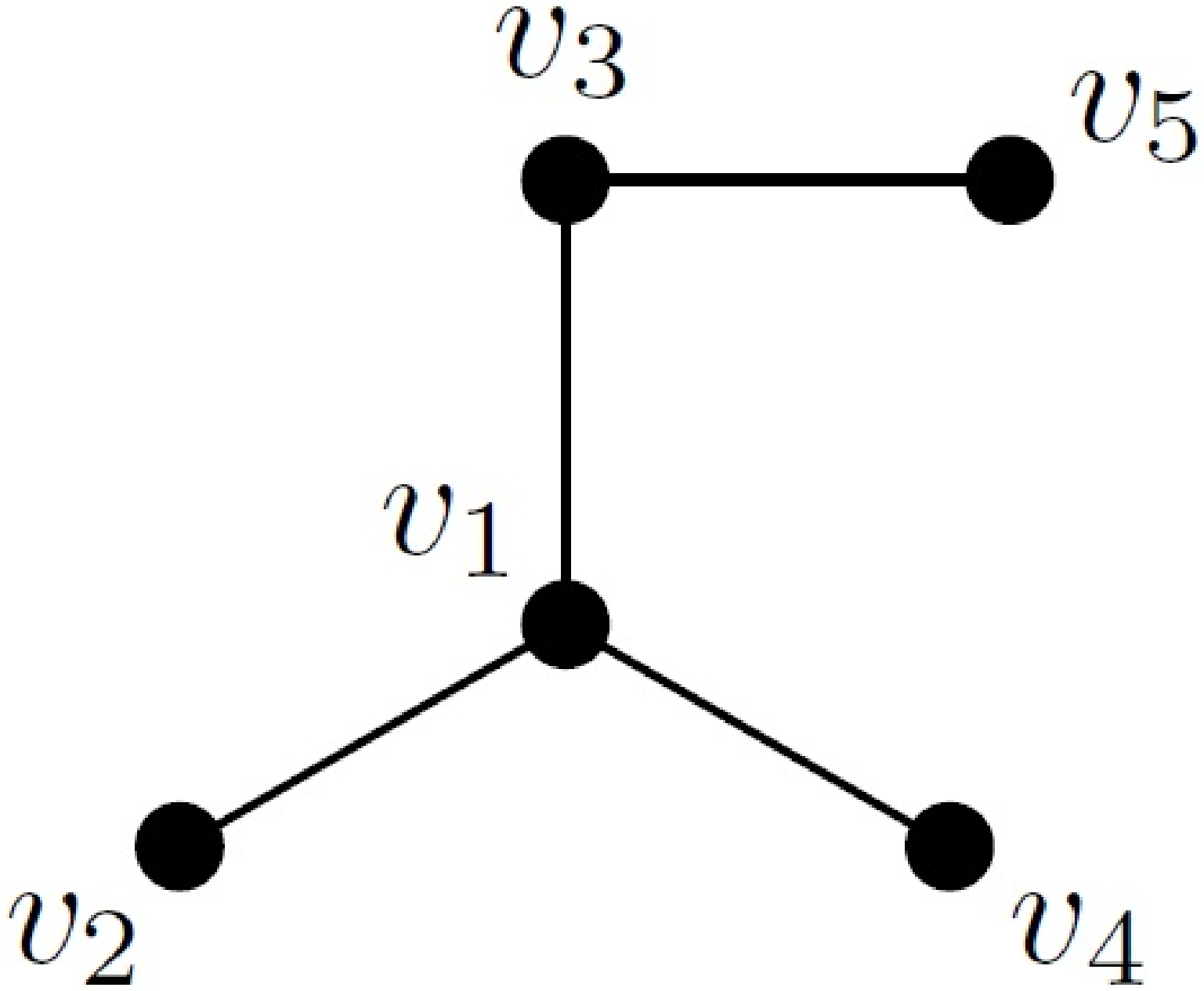}
\end{center}
\vspace{-20pt}
  \caption{}
  \label{fig:grasu}
\end{figure}

There exist some differences between GCP and ECP that make the latter harder to solve.
It is known that the chromatic number of a graph is greater than or equal to the chromatic number of any of its induced subgraphs.
Unfortunately, in the case of ECP, this property does not hold. For instance, if $G$ is the graph shown in Figure \ref{fig:grasu},
by deleting $v_5$ from $G$, $\chi_{eq}(G)$ increases from 2 to 3.
%Another difference between both problems is that, in ECP, a graph admiting a $k$-eqcol may not admit a $(k+1)$-eqcol.
%For instance, if $G = K_{3,3}$, \ie the complete bipartite graph with partitions of size 3, then $G$ admits a 2-eqcol
%but does not admit a 3-eqcol. Those graphs $G$ admitting an $r$-eqcol for each $r \in \{\chi_{eq}(G), \ldots, n\}$,
%where $n = |V|$, are called \emph{monotones}. For instance, trees are monotone graphs \cite{EQTREE}.

As far as we know, there are few approximate and exact algorithms available in the literature related to ECP.

It was proved that, for any graph $G$, $\Delta(G) + 1$ is an upper bound of $\chi_{eq}(G)$ \cite{ERDOS}, where $\Delta(G)$ is the maximum degree of vertices
in $G$. Based on this fact, a polynomial time algorithm for obtaining a $k$-eqcol of a graph $G$ with $k \geq \Delta(G)+1$ is described in \cite{MARCELO}.

Two constructive heuristics called \textsc{Naive} and \textsc{SubGraph} are given in \cite{KUBALE} to generate greedily an equitable coloring of a graph. There also exist heuristic algorithms for constructing colorings that are ``nearly'' equitable
\cite{BRELAZEQUIT,OTRO}, making emphasis on achieving a small difference between the sizes of the biggest
class and the smallest one, although the equity constraint still might be violated.

The authors of \cite{BYCBRA} propose a tabu search heuristic to initialize an exact algorithm that solves ECP
via Integer Linear Programming (ILP) techniques. Other exact algorithms for solving ECP are given in \cite{PAPERDAM} and \cite{EQDSATUR}.
The first one also uses IPL techinques and the second one is based on a DSATUR enumeration scheme.
% Also, in \cite{PAPERDAM}, an heuristic that returns a lower bound of the equitable chromatic number is described.

In this work, we propose a new heuristic based on the dynamic \textsc{TabuCol} version of Galinier and Hao \cite{GALINIER}, one of the best tabu
search algorithms for GCP \cite{SURVEY}. Then, computational experiments are carried out in order to find the best combination of parameters
involved in the dynamic tenure of our heuristic and to show the good performance of it over known benchmark instances.

The paper is organized as follows. In Section \ref{SGCP}, we present \textsc{TabuCol} and the dynamic variant of Galinier and Hao.
In Section \ref{SECP}, we give our variant for ECP which we call \textsc{TabuEqCol}. Finally, in Section \ref{SCOMPU} we report
computational experiences and conclusions.

\section{\textsc{TabuCol} and its variants} \label{SGCP}

\emph{Tabu search} is a metaheuristic method proposed by Glover \cite{GLOVER} that guides a local search algorithm equipped with additional mechanisms that prevent from visiting a solution twice and getting stuck in a local optimum. 

Let $S$ be the solution space of the problem and $f : S \rightarrow \mathbb{R}$ be the objective function. The goal is to obtain a
solution $s \in S$ such that $f(s)$ is minimum.

For each solution $s \in S$, consider a \emph{neighborhood} $N(s) \subset S$ with two desirable (but not exclusionary)
properties: 1) two solutions $s$ and $s'$ are neighbors when it is easy (from the computational point of view) to
obtain $s'$ from $s$, and to obtain $f(s')$ from $f(s)$ (for instance, in constant time), and 2) for any $s, s' \in S$,
there exists a path $s = s_1, s_2, \ldots, s_m = s'$ such that $s_{i+1} \in N(s_i)$ for $i = 1,\ldots,m-1$.

In general, neighbor solutions are very similar in some sense, and the difference between them can be seen as \emph{features} that
both solutions do not share. Consider a set of \emph{features} $P$ and a set $R \subset S \times P$ such that
$(s, p) \in R$ if solution $s$ presents a feature $p$.

Starting from an initial solution $s_0 \in S$, tabu search consists of generating a sequence of solutions
$s_1, s_2, \ldots$ such that $s_{i+1} = \textrm{arg min}_{s \in N'(s_i)} f(s)$, where $N'(s_i)$ is a subset of $N(s_i)$
described below. In each iteration of this algorithm, a \emph{movement} from $s_i$ to $s_{i+1}$ is performed and
some feature of $s_i$ is stored in a \emph{tabu list} $L \subset P$. This list indicates whether a movement is allowed or
forbidden: a solution $s$ can be reached in the future only if $s$ does not present any feature from $L$
(this rule avoids from visiting a solution previously visited), except when $s$ is better than the best solution found so
far. This exception is called \emph{aspiration} and the aspiration criterion is usually to check if the objective value
of $s$ is less than the value of currently-known best solution. Now, the set of allowed movements from $s_i$, $N'(s_i)$, is defined as
\[ N'(s) = \{ s' \in N(s) : f(s') < f(s^*) ~\lor~ (s',p) \notin R ~~ \forall~p \in L \}, \]
where $s^*$ is the best solution found so far.

However, after several iterations, old features are no longer needed and it is better to remove them from the tabu list.
This mechanism is usually implemented by assigning a ``time of live'' to each feature of the tabu list.
Consider $live : L \rightarrow \mathbb{Z}$ and let $live(p)$ be the number of remaining iterations that $p$ belongs to $L$.
When a new feature $p$ is inserted into $L$, $live(p)$ is assigned a value referred to as \emph{tabu tenure} $t$.
Then, in each iteration, the value of $live(p)$ is decreased by one unit until it reachs zero and $p$ is removed from $L$.
Above, we sketch a generic tabu search algorithm.

\begin{algorithm}
\DontPrintSemicolon
\KwData{initial solution $s_0$}
\SetKwData{undefined}{undefined}
\KwResult{best solution found $s^*$}
\Begin{
  $L \leftarrow \varnothing$ \;
  $s, s^* \leftarrow s_0$ \;
  \While{stopping criterion is not met}{
    \For{$p \in L$}{
      $live(p) \leftarrow live(p) - 1$ \;
      if $live(p) = 0$ then $L \leftarrow L \backslash \{p\}$ \;
    }
    $N'(s) \leftarrow \{ s' \in N(s) : f(s') < f(s^*) ~\lor~ (s',p) \notin R ~~ \forall~p \in L \}$ \;
    choose a feature $p \in P$ such that $(s,p) \in R$ \;
    $L \leftarrow L \cup \{p\}$ \;
    $live(p) \leftarrow t$ \;
    $s \leftarrow \textrm{arg min}_{s' \in N'(s)} f(s')$ \;
    if $f(s) < f(s^*)$ then $s^* \leftarrow s$ \;
  }
}
\caption{\textsc{TabuSearch}}
\end{algorithm}

In order to implement a tabu search algorithm, some decisions must be taken: neighborhood of a solution, features of a solution,
stopping criterion, how to choose the feature $p$ to be stored in the tabu list and how to compute the tabu tenure $t$.
In particular, the value of tabu tenure directly impacts \emph{diversification} of the algorithm.
A tabu search with low tenures behaves as a standard local search, where it frequently get trapped in local minima.
On the other hand, a tabu search with high tenures tends to wander across solution space without converging towards the optimal
solution.\\

\textsc{TabuCol}, the first tabu search algorithm designed for solving GCP, was proposed by Hertz and de Werra \cite{TABUCOL}.
For a given graph $G = (V, E)$ and number $k \in \{1, \ldots, n\}$, where $n = |V|$, the goal of this algorithm is to find a
$k$-coloring of $G$.
In order to obtain a coloring that uses as few colors as possible, it is usual to embed \textsc{TabuCol} in a routine that,
once a $k$-coloring is found, the algorithm can be restarted with $k \leftarrow k - 1$ and so on, until some criterion is met.
Details of \textsc{TabuCol} are given below:
\begin{itemize}
\item \emph{Search space and objective function}. A solution $s$ is a partition $(V_1, V_2, \ldots, V_k)$ of the set of vertices.
Let $E(V_i)$ be the set of edges of $G$ with both endpoints in $V_i$. The objective function is defined as
\[ f(s) = \sum_{i=1}^k |E(V_i)|. \]
Clearly, $s$ is a $k$-coloring if and only if $f(s) = 0$.
\item \emph{Stopping criterion}. The algorithm stops when $f(s) = 0$ or when a maximum number of iterations is reached.
Sometimes, a time limit is imposed.
\item \emph{Initial solution}. It is generated randomly. A suitable procedure given in \cite{REACTTABU} is the following.
Start with empty sets $V_1, V_2, \ldots, V_k$ and, at each step, choose a non-considered vertex $v$ randomly and put it into
$V_i$ with the smallest possible $i$ such that $E(V_i)$ is not incremented. If it is not possible, choose a random number
$j \in \{1, \ldots, k\}$ and put $v$ into $V_j$.
\item \emph{Set of features}. It is $P = V \times \{1, \ldots, k\}$. A solution $s$ presents a feature $(v, i)$ if and only if
$v \in V_i$, \ie if $v$ is assigned color $i$.
\item \emph{Neighborhood of a solution}. Let $C(s)$ be the \emph{set of conflicting vertices} of a solution $s$, \ie
\[ C(s) = \{ v \in V : \textrm{$v$ is incident in some edge of $E(V_1) \cup E(V_2) \cup \ldots \cup E(V_k)$} \}. \]
From a solution $s = (V_1, V_2, \ldots, V_k)$, a neighbor $s'  = (V'_1, V'_2, \ldots, V'_k)$ is generated as follows. Choose
a conflicting vertex $v \in C(s)$. Let $i$ be the color of $v$ in $s$. Next, choose a color $j \in \{1, \ldots, k\} \backslash \{i\}$
and set
\[ V'_j = V_j \cup \{ v\},~~  V'_i = V_i \backslash \{v\},~~  V'_l = V_l ~~ \forall~l \in \{1,\ldots,k\} \backslash \{i, j\}. \]
In other words, $s'$ is a copy of $s$ except that $v$ is moved from class color $V_i$ to $V_j$. We denote such operation with
$s' = s(i \xrightarrow{v} j)$. Note that objective value can be computed in linear time from $f(s)$:
\[ f(s') = f(s) + |\{ vw \in E : w \in V_j \}| - |\{ vw \in E : w \in V_i \}|. \]
Note also that searching all the neighbors of $s$ requires exploring $(k-1)|C(s)|$ solutions.
Original \textsc{TabuCol} only explores a random subset of $N(s)$ while newer versions explore $N(s)$ completely.
\item \emph{Selection of feature to add in the tabu list}. Once a movement from $s$ to $s(i \xrightarrow{v} j)$ is performed,
$p=(v,i)$ is stored on tabu list and $live(p)$ is set to a fixed tabu tenure $t = 7$.
\end{itemize}

Later, Galinier and Hao \cite{GALINIER} improved \textsc{TabuCol} by using a dynamic tabu tenure that depends on the quality of
the current solution, encouraging diversification of the search when solution is far from optimal.
They proposed to assign a tenure of $t = \alpha|C(s)| + Random(\beta)$ where $Random(\beta)$ returns an integer randomly chosen from $\{0, \ldots, \beta-1\}$ with uniform distribution.
Based on experimentation, they suggest to use $\alpha = 0.6$ and $\beta = 10$.
Other variants of \textsc{TabuCol} are discussed in \cite{SURVEY,REACTTABU}.

\section{\textsc{TabuEqCol}: A tabu search for ECP} \label{SECP}

In this section, we present a new tabu search algorthm for ECP based on \textsc{TabuCol} with
dynamic tabu tenure, which we call \textsc{TabuEqCol}.

Given a graph $G = (V, E)$ and a number $k \in \{1,\ldots,n\}$, where $n = |V|$, the goal of \textsc{TabuEqcol}
is to find a $k$-eqcol of $G$.

Solution space consists of partitions of $V$ into $k$ sets $V_1, V_2, \ldots, V_k$ such that they satisfy
the equity constraint, \ie for any pair of classes $V_i$ and $V_j$, $\bigl| |V_i|-|V_j| \bigr| \leq 1$.
Objective function $f$ is the same as in \textsc{TabuCol}, so any solution $s$ such
that $f(s) = 0$ is indeed an equitable coloring.
Also, set of features $P$ is the same as in \textsc{TabuCol}.

Stopping criterion depends on the experiment carried out. Usually, a time limit is imposed.\\

Let $s \in S$. Denote $W^+(s) = \{ i : |V_i| = \lfloor n/k \rfloor+1\}$ and
$W^-(s) = \{ i : |V_i| = \lfloor n/k \rfloor\}$, where $V_i$ are the color classes of $s$. Since $s$ satisfies the equity
constraint, we have that $W^+(s)$ and $W^-(s)$ determine a partition of $\{1,\ldots,k\}$ and, in particular, $|W^+(s)| = r$ where
$r = n - k\lfloor n / k \rfloor$.  From now on, we just write $W^+$ and $W^-$. These sets will be useful in the development
of the algorithm. %In particular, for each solution $s$ in memory it is convenient to keep a copy of $W^+$.

We propose two greedy procedures for generating initial solution $s_0$.\\

\noindent \emph{Procedure 1}. Start with empty sets $V_1, V_2, \ldots, V_k$ and an
integer $\tilde{r} \leftarrow 0$ (this value will have the cardinal of $W^+$). At each step,
define set $I = \{ i : |V_i|  \leq M-1 \}$, where $M$ is the maximum allowable size of a class:
\[ M = \begin{cases}
\lfloor n/k \rfloor + 1, & \textrm{if}~~\tilde{r} < r \\
\lfloor n/k \rfloor,     & \textrm{if}~~\tilde{r} = r
\end{cases} \]
(once we already have $r$ class of size $\lfloor n/k \rfloor + 1$, the size of the remaining classes must not exceed
$\lfloor n/k \rfloor$). Then, choose a non-considered vertex $v$ randomly and put it into a class $V_i$ such that $i \in I$ is the smallest possible and
$E(V_i)$ is not incremented. If it is not possible, $i$ is chosen ramdonly from $I$.
To keep $\tilde{r}$ up to date, each time a vertex is added to a set $V_i$ such that $|V_i| = \lfloor n/k \rfloor$,
$\tilde{r}$ is incremented by one unit.\\

The previous procedure works fine for generating initial solutions from scratch. However, at this point it is common to know a
$(k+1)$-eqcol (\ie in the cases where we previously ran tabu search with $k+1$ and reached an equitable coloring) and
we can exploit this coloring in order to improve the quality of the initial solution as follows.\\

\noindent \emph{Procedure 2}. Let $\mathfrak{p}:\{1,\ldots,k+1\} \rightarrow \{1,\ldots,k+1\}$ be a bijective function
(\ie a random permutation) and let $V^*_1, V^*_2, \ldots, V^*_k, V^*_{k+1}$ be the color classes of the known $(k+1)$-eqcol.
Set $V_i = V^*_{\mathfrak{p}(i)}$ for all $i \in \{1, \ldots, k\}$, and $\tilde{r} = |W^+|$. Then, run Procedure 1 to assign a
color to the remaining vertices which are those belonging to $V^*_{\mathfrak{p}(k+1)}$.\\

Regarding neighborhood of a solution $s \in S$ notice that, if $n$ does not divide $k$, $W^+ \neq \varnothing$ and it is possible to move a vertex from a class of $W^+$ to $W^-$, keeping equity. That is, for all $v \in \cup_{i \in W^+} V_i$ and all $j \in W^-$,
we have $s(i \xrightarrow{v} j) \in S$.
However, the number of allowed movements is rather limited when $r$ is very low (for instance, $r = 1$) or very high ($r = k-1$),
so we need to add supplementary movements.
Swapping the colors of two vertices simultaneously seems to work fine and as well can be used when $n$ divides $k$.

From a solution $s = (V_1, V_2, \ldots, V_k)$, a neighbor $s'  = (V'_1, V'_2, \ldots, V'_k)$ is generated with two schemes:
\begin{itemize}
\item \emph{1-move} (only applicable when $n$ does not divide $k$). Choose a conflicting vertex
$v \in C(s) \cap (\cup_{i \in W^+} V_i)$. Let $i$ be the color of $v$ in $s$. Next, choose a color $j \in W^-$. We have
$s' = s(i \xrightarrow{v} j)$. Searching all the neighbors of $s$ with 
this scheme requires exploring $(k-r)|C(s) \cap (\cup_{i \in W^+} V_i)|$ solutions.
\item \emph{2-exchange}. Choose a conflicting vertex $v \in C(s)$. Let $i$ be the color of $v$ in $s$. Next, choose another vertex
$u$ such that either $i < j$ or $u \notin C(s)$, where $j$ is the color of $u$ in $s$ (the condition imposed to $u$ prevents from
evaluating 2-exchange on $u$ and $v$ twice). Then, set
\[ V'_j = (V_j \backslash \{u\}) \cup \{v\},~~  V'_i = (V_i \backslash \{v\}) \cup \{u\},~~  V'_l = V_l ~~ \forall~l \in \{1,\ldots,k\} \backslash \{i, j\}. \]
Note that objective value can be computed in linear time from $f(s)$:
\begin{multline*}
  f(s') = f(s) + |\{ uw \in E : w \in V_i\backslash\{v\} \}| - |\{ uw \in E : w \in V_j \}| \\
  + |\{ vw \in E : w \in V_j\backslash\{u\} \}| - |\{ vw \in E : w \in V_i \}|.
\end{multline*}
Searching all the neighbors of $s$ with this scheme requires exploring a quadratic number of solutions.
\end{itemize}

Now, let $s'$ be the next solution in the sucession; $s'$ is obtained by applying either 1-move or 2-exchange to $s$, where vertex
$v \in V_i$ in $s$ and $v \notin V'_i$ in $s'$. In both schemes, $p=(v,i)$ is stored on tabu list and $live(p)$ is set to a dynamic tabu tenure
$t = \alpha|C(s)| + Random(\beta)$ where $\alpha$ and $\beta$ are parameters to be determined empirically. This is one of the purposes of the next section.

\section{Computational experiments and conclusions} \label{SCOMPU}

This section is devoted to perform and analyze computational experiments. They were carried out on an Intel i5 CPU 750@2.67Ghz with Ubuntu Linux O.S.
and Intel C++ Compiler. We considered graphs from \cite{DIMACS}, which are benchmark instances difficult to color.

First, we  test different combinations of values for parameters $\alpha$ and $\beta$ from the dynamic tabu tenure in order to determine the
combination that makes \textsc{TabuEqCol} perform better.
Then, we report the behaviour of \textsc{TabuEqCol} over known instances by using the best combination previously found.
We also compare its performance against tabu search algorithm given in \cite{BYCBRA}.\\

\noindent \emph{Tuning parameters}\\

We run \textsc{TabuEqCol} over 16 instances with a predetermined value of $k$ and an initial solution $s_0$ generated
with Procedure 1. The same initial solution is used in all executions of \textsc{TabuEqCol} for the same instance.

Results are reported in Table \ref{TABLE1}. First column is the name of the graph $G$. Second and third columns
are the number of vertices and edges of $G$. Fourth and fifth columns are known lower and upper bound of $\chi_{eq}(G)$ (obtained by other means).
The remaining columns are the time elapsed in seconds by the execution of \textsc{TabuEqCol} when
a $k$-eqcol is found within the term of 1 hour, for each combination. In the case \textsc{TabuEqCol} is not able to find a $k$-eqcol,
$f(s^*)$ is displayed between braces where $s^*$ is the best solution found.
Three last rows indicate the sum of objective function $f(s^*)$ over non-solved instances, percentage of instances \textsc{TabuEqCol}
solved successfully and the average time elapsed for these instances to be solved. 

For the sake of simplicity, we refer to each combination with a capital letter.\\

Note that combination D has the least average time, however it has solved less instances than other combinations
and the sum of objective values is also worse. We discard A, B, C, D, E and H with this criterion. By comparing the three remaining combinations,
we have that G is faster than the other two. Even if we restrict the comparison to those 11 instances the 3 combinations solve simultaneously,
we have 807 seconds for F, 562 seconds for G and 730 seconds for I, so G is still better.
 
%Note that combinations F and G have solved more instances than the others. In terms of time, combination G
%is 34\% faster than F. If we restrict the comparison to those 11 instances both F and G solves simultaneously, we have 807
%seconds for F and 562 seconds for G, so G is still better than F.
%On the other hand, combination D performs 37\% better than G in terms of time, but if we compare average time over those 9 instances both
%D and G solve simultaneously, the gap is reduced to 28\% since we have 456 seconds for D and 631 seconds for G.
%We now see which combination obtains the best value of $f(s^*)$: G has not been able to solve \texttt{flat300\_28\_0} but
%$f(s^*) = 1$, meanwhile D has not solved \texttt{DSJR500.1c}, \texttt{abb313GPIA} nor \texttt{wap01a} and the values of $f(s^*)$ are
%worse; also for instance \texttt{DSJR500.5}, G has reached a better value of $f(s^*)$ than D.

We consider combination G ($\alpha = 0.9$ and $\beta = 5$) for \textsc{TabuEqCol}.\\

\noindent \emph{Testing tabu search heuristic}\\

For each instance, the following process is performed. First, execute \textsc{Naive} algorithm (described in \cite{KUBALE}) in order
to find an initial equitable coloring $c$ of the current instance. Suppose that $k+1$ is the number of colors of $c$. Then, obtain an
initial solution $s_0$ of $k$ color classes generated from $c$ with Procedure 2, and run \textsc{TabuEqcol} with parameters $\alpha=0.9$ and $\beta=5$.
If a $k$-eqcol is found, start over the process with $k-1$ color classes by running Procedure 2 and \textsc{TabuEqcol} again.
This process is repeated until 1 hour is elapsed or a $\underline{\chi_{eq}}$-eqcol is reached, and the best coloring found so far
is returned.

In Table \ref{TABLE2} we report results over 76 benchmark instances with at least 50
vertices (75 from \cite{DIMACS} and one Kneser graph used in \cite{BYCBRA}). First 5 columns have the name of the graph $G$, number of vertices
and edges, and best known lower and upper bound of $\chi_{eq}(G)$. Sixth column displays the number
of colors of the initial equitable coloring $c$. Seventh and eighth columns display the value $k$ of the best $k$-eqcol found after 30 seconds
of execution of our algorithm and the time elapsed in seconds until such $k$-eqcol is reached. If the coloring is optimal, $k$ is displayed in boldface. Next two columns show the same information
after 1 hour of execution, but if the best coloring is found within the first 30 seconds, these columns are left empty.

Time spent by \textsc{Naive} is not considered in the computation. However, \textsc{Naive} rarely spent more than 1 sec. (and never more than 4 sec.).

Last two columns show the same information for the tabu search described in \cite{BYCBRA}. If such information is not available, these columns
are left empty. We recall that the values provided in \cite{BYCBRA} were computed on a different platform (1.8 Ghz AMD-Athlon with Linux and GNU C++ compiler).\\

Note that our approach reachs optimality in 29 instances and a gap of one unit between $\underline{\chi_{eq}}$
and the best solution in 7 instances. In other words, it reachs a gap of at most one unit in roughly a half of the evaluated instances. Note also that
\textsc{TabuEqcol} improves the initial solution given by \textsc{Naive} in most cases (precisely, 63 instances).

On those instances the value of the best solution given by tabu search of \cite{BYCBRA} is known, our algorithm gives the same value or a
better one. Despite the difference between platforms, it seems that our approach also runs faster.

An interesting fact is that each execution of \textsc{TabuEqCol} needs no more than 500000 iterations to reach the
best value since the largest number of iterations performed was 493204 and took place when \textsc{TabuEqcol} found a 18-eqcol
of \texttt{DSJC125.5}.

In the same sense, \textsc{TabuEqCol} needs no more than 30000 iterations in each execution and the overall process needs
no more than 30 seconds to reach the best value on 56 instances; justly those ones such that columns 9 and 10 are empty.
On these instances, the largest number of
iterations performed was 28791 and took place when \textsc{TabuEqcol} found a 10-eqcol of \texttt{queen9\_9}.\\

\noindent \emph{Conclusion}\\

The Equitable Coloring Problem is a variation of the Graph Coloring Problem that naturally arises from several applications where the cardinalities of color
classes must be balanced. Just like Graph Coloring, the need to solve applications associated to this new NP-Hard problem justifies the development of exact
and approximate algorithms.
On large instances, known exact algorithms are unable to address them and heuristics such as \textsc{Naive} delivers poor solutions.
Our tabu search heuristic based on \textsc{TabuCol} has shown to improve these solutions and
%It is only natural to adapt algorithm to ECP that actually performs well on GCP. In this sense, \textsc{TabuCol} is a good
%candidate. In the words of Galinier and Hertz, ``... almost all efficient heuristic algorithms for graph coloring use a local search, and many of them are
%based on a tabu search. In particular, \textsc{TabuCol} is very popular, the reason being probably that it is a very simple algorithm that is easy to
%implement. In addition, \textsc{TabuCol} offers a good compromise between solution quality and computational effort.'' \cite{SURVEY}.
presented a fairly good performance, even if a limit of 30 seconds is imposed. In addition, an iteration limit of 30000
(for a time limit of 30 seconds) and 500000 (for a time limit of 1 hour) can be imposed in order to save time.\\
%In the first case, \textsc{TabuEqCol}
%can be used as part of a major algorithm (such as Branch and Cut) to bring initial colorings. In the second case, it can be used as a stand alone tool
%to generate good solutions.

\noindent \textbf{Acknowledgements}. This work is partially supported by grants
UBACYT 20020100100666, PICT 2010-304, PICT 2011-817, PID-UNR ING416 and PIP-CONICET 241.

\begin{sidewaystable}
\centering
\small
\begin{tabular}{ccccc|ccc|ccc|ccc}
 & & & & & \multicolumn{3}{|c}{$\alpha=0.3$} & \multicolumn{3}{|c}{$\alpha=0.6$} & \multicolumn{3}{|c}{$\alpha=0.9$} \\
Instance & $|V|$ & $|E|$ & $\underline{\chi_{eq}}$ & $k$ & $\beta=5$ & $\beta=10$ & $\beta=15$ & $\beta=5$ & $\beta=10$ & $\beta=15$ &
 $\beta=5$ & $\beta=10$ & $\beta=15$ \\
 & & & & & A & B & C & D & E & F & G & H & I \\
\hline
\texttt{DSJR500.1} & 500 & 3555 & 12 & 12 & \{3\} & 1 & 1 & 1 & 1 & 1 & 1 & 1 & 1 \\
\texttt{DSJR500.5} & 500 & 58862 & 120 & 131 & \{14\} & \{3\} & \{1\} & \{8\} & \{3\} & 3242 & \{5\} & \{3\} & \{1\} \\
\texttt{DSJR500.1c} & 500 & 121275 & 126 & 195 & \{4\} & \{1\} & 427 & \{3\} & 78 & 747 & 66 & 8 & 11 \\
\texttt{DSJC500.1} & 500 & 12458 & 5 & 13 & \{2\} & 55 & 41 & 38 & 63 & 47 & 39 & 83 & 57 \\
\texttt{DSJC500.5} & 500 & 62624 & 13 & 62 & 61 & 530 & \{1\} & \{1\} & \{1\} & \{2\} & \{1\} & \{2\} & \{1\} \\
\texttt{DSJC500.9} & 500 & 112437 & 101 & 148 & \{1\} & 106 & 104 & 94 & 91 & 80 & 100 & 90 & 121 \\
\texttt{DSJC1000.1} & 1000 & 49629 & 5 & 22 & 767 & 411 & 509 & 551 & 423 & 858 & 710 & 691 & 1059 \\
\texttt{DSJC1000.5} & 1000 & 249826 & 15 & 112 & 543 & 968 & 623 & 518 & 999 & \{2\} & 1853 & \{2\} & \{1\} \\
\texttt{DSJC1000.9} & 1000 & 449449 & 126 & 268 & 1850 & 1751 & 1822 & 1926 & 1725 & 1250 & 1808 & 1723 & 983 \\
\texttt{inithx.i.1} & 864 & 18707 & 54 & 54 & \{8\} & \{8\} & \{8\} & \{8\} & \{7\} & \{7\} & \{8\} & \{8\} & \{7\} \\
\texttt{latin\_square\_10} & 900 & 307350 & 90 & 131 & 1182 & 1080 & 1013 & 796 & 782 & 946 & 895 & 1298 & 778 \\
\texttt{flat300\_28\_0} & 300 & 21695 & 11 & 37 & 238 & \{1\} & \{1\} & 143 & \{1\} & \{1\} & \{1\} & \{2\} & \{2\} \\
\texttt{flat1000\_76\_0} & 1000 & 246708 & 14 & 112 & 228 & 548 & 1255 & 154 & 600 & 1681 & 245 & 780 & 3298 \\
\texttt{abb313GPIA} & 1557 & 53356 & 8 & 9 & \{27\} & \{44\} & \{15\} & \{2\} & \{10\} & 2801 & 1796 & \{1\} & 1304 \\
\texttt{qg.order40} & 1600 & 62400 & 40 & 40 & 26 & 31 & 17 & 25 & 26 & 20 & 24 & 25 & 26 \\
\texttt{wap01a} & 2368 & 110871 & 41 & 47 & \{21\} & 477 & 501 & \{6\} & 451 & 446 & 499 & 744 & 397 \\
\hline
\multicolumn{5}{c|}{Sum of objective values} & 80 & 57 & 26 & 28 & 22 & 12 & 15 & 18 & 12 \\
\multicolumn{5}{c|}{Success} & 50\% & 69\% & 69\% & 63\% & 69\% & 75\% & 75\% & 63\% & 69\% \\
\multicolumn{5}{c|}{Average Time} & 612 & 542 & 574 & 425 & 476 & 1010 & 670 & 544 & 730
\end{tabular}
\caption{Execution of \textsc{TabuEqCol} with different combination of values}
\label{TABLE1}
\end{sidewaystable}

\begin{table}
\centering
\tiny
\begin{tabular}{ccccc|c|cc|cc|cc}
 & & & & & & \multicolumn{2}{|c}{$\leq$ 30 sec.} & \multicolumn{2}{|c}{$\leq$ 1 hour} & \multicolumn{2}{|c}{\cite{BYCBRA}} \\
Instance & $|V|$ & $|E|$ & $\underline{\chi_{eq}}$ & $\overline{\chi_{eq}}$ & \textsc{Naive} & $k$ & Time & $k$ & Time & $k$ & Time \\
% &  &  &  &  & &  &  &  &  &  &  \\
\hline
\texttt{miles750} & 128 & 2113 & 31 & 31 & 33 & \textbf{31} & 0.0 & & & 35 & 13 \\
\texttt{miles1000} & 128 & 3216 & 42 & 42 & 47 & 43 & 0.1 & & & 49 & 13 \\
\texttt{miles1500} & 128 & 5198 & 73 & 73 & 74 & \textbf{73} & 0.0 & & & 77 & 13 \\
\texttt{zeroin.i.1} & 211 & 4100 & 49 & 49 & 51 & 51 & 0.0 & & & 74 & 22 \\
\texttt{zeroin.i.2} & 211 & 3541 & 36 & 36 & 51 & 51 & 0.0 & & & 95 & 22 \\
\texttt{zeroin.i.3} & 206 & 3540 & 36 & 36 & 49 & 49 & 0.0 & & & 97 & 21 \\
\texttt{queen8\_8} & 64 & 728 & 9 & 9 & 18 & \textbf{9} & 1.2 & & & 10 & 7 \\
\texttt{jean} & 80 & 254 & 10 & 10 & 10 & \textbf{10} & 0.0 & & & \textbf{10} & 3 \\
\texttt{anna} & 138 & 493 & 11 & 11 & 11 & \textbf{11} & 0.0 & & & 13 & 14 \\
\texttt{david} & 87 & 406 & 30 & 30 & 40 & \textbf{30} & 0.0 & & & \textbf{30} & 9 \\
\texttt{games120} & 120 & 638 & 9 & 9 & 9 & \textbf{9} & 0.0 & & & 11 & 6 \\
\texttt{kneser9\_4} & 126 & 315 & 3 & 3 & 4 & \textbf{3} & 0.0 & & & 6 & 2 \\
\texttt{2-FullIns\_3} & 52 & 201 & 5 & 5 & 9 & \textbf{5} & 0.0 & & & 8 & 1 \\
\texttt{3-FullIns\_3} & 80 & 346 & 6 & 6 & 7 & \textbf{6} & 0.0 & & & 9 & 2 \\
\texttt{4-FullIns\_3} & 114 & 541 & 7 & 7 & 12 & \textbf{7} & 0.1 & & & 11 & 5 \\
\texttt{5-FullIns\_3} & 154 & 792 & 8 & 8 & 9 & \textbf{8} & 0.0 & & & 13 & 8 \\
\texttt{2-FullIns\_5} & 852 & 12201 & 4 & 7 & 15 & 7 & 2.5 & & & & \\
\texttt{3-FullIns\_5} & 2030 & 33751 & 5 & 8 & 13 & 8 & 25 & & & & \\
\texttt{4-FullIns\_4} & 690 & 6650 & 6 & 8 & 14 & 8 & 0.4 & & & & \\
\texttt{4-FullIns\_5} & 4146 & 77305 & 6 & 9 & 21 & 14 & 20 & 9 & 254 & & \\
\texttt{1-Insertions\_6} & 607 & 6337 & 3 & 7 & 14 & 7 & 0.2 & & & & \\
\texttt{2-Insertions\_5} & 597 & 3936 & 3 & 6 & 6 & 6 & 0.0 & & & & \\
\texttt{3-Insertions\_5} & 1406 & 9695 & 3 & 6 & 8 & 6 & 1.2 & & & & \\
\texttt{homer} & 561 & 1628 & 13 & 13 & 13 & \textbf{13} & 0.0 & & & & \\
\texttt{huck} & 74 & 301 & 11 & 11 & 11 & \textbf{11} & 0.0 & & & & \\
\texttt{latin\_square\_10} & 900 & 307350 & 90 & 130 & 460 & 169 & 30 & 130 & 1301 & & \\
\texttt{DSJC125.1} & 125 & 736 & 5 & 5 & 8 & \textbf{5} & 0.8 & & & & \\
\texttt{DSJC125.5} & 125 & 3891 & 9 & 18 & 27 & 19 & 0.1 & 18 & 788 & & \\
\texttt{DSJC125.9} & 125 & 6961 & 42 & 45 & 66 & 45 & 0.4 & & & & \\
\texttt{DSJC250.1} & 250 & 3218 & 4 & 8 & 13 & 9 & 0.1 & 8 & 32 & & \\
\texttt{DSJC250.5} & 250 & 15668 & 11 & 32 & 65 & 33 & 7.2 & 32 & 69 & & \\
\texttt{DSJC250.9} & 250 & 27897 & 63 & 83 & 136 & 83 & 1.2 & & & & \\
\texttt{DSJR500.1} & 500 & 3555 & 12 & 12 & 12 & \textbf{12} & 0.0 & & & & \\
\texttt{DSJR500.5} & 500 & 58862 & 120 & 131 & 135 & 133 & 0.1 & & & & \\
\texttt{DSJR500.1c} & 500 & 121275 & 126 & 195 & 349 & 257 & 0.3 & & & & \\
\texttt{DSJC500.1} & 500 & 12458 & 5 & 13 & 23 & 14 & 3.5 & 13 & 33 & & \\
\texttt{DSJC500.5} & 500 & 62624 & 13 & 62 & 128 & 63 & 11 & & & & \\
\texttt{DSJC500.9} & 500 & 112437 & 101 & 148 & 284 & 182 & 0.7 & & & & \\
\texttt{DSJC1000.1} & 1000 & 49629 & 5 & 22 & 38 & 26 & 26 & 22 & 500 & & \\
\texttt{DSJC1000.5} & 1000 & 249826 & 15 & 112 & 265 & 128 & 27 & 112 & 2261 & & \\
\texttt{DSJC1000.9} & 1000 & 449449 & 126 & 268 & 575 & 329 & 20 & & & & \\
\texttt{flat300\_20\_0} & 300 & 21375 & 11 & 34 & 81 & 38 & 9.2 & 34 & 463 & & \\
\texttt{flat300\_28\_0} & 300 & 21695 & 11 & 36 & 65 & 39 & 3.3 & 36 & 3222 & & \\
\texttt{flat1000\_76\_0} & 1000 & 246708 & 14 & 112 & 223 & 127 & 24 & 112 & 1572 & & \\
\texttt{fpsol2.i.1} & 496 & 11654 & 65 & 65 & 85 & 78 & 0.1 & & & & \\
\texttt{fpsol2.i.2} & 451 & 8691 & 47 & 47 & 62 & 60 & 0.0 & & & & \\
\texttt{fpsol2.i.3} & 425 & 8688 & 55 & 55 & 80 & 79 & 0.0 & & & & \\
\texttt{inithx.i.1} & 864 & 18707 & 54 & 54 & 70 & 66 & 0.1 & & & & \\
\texttt{inithx.i.2} & 645 & 13979 & 30 & 93 & 158 & 93 & 7.2 & & & & \\
%\texttt{kneser11\_5} & 462 & 1386 & 3 & 3 & 4 & 4 & 0.0 & & & & \\
\texttt{le450\_15b} & 450 & 8169 & 15 & 15 & 17 & 16 & 0.3 & \textbf{15} & 107 & & \\
\texttt{le450\_15d} & 450 & 16750 & 15 & 16 & 30 & 22 & 9.6 & 16 & 599 & & \\
\texttt{le450\_25b} & 450 & 8263 & 25 & 25 & 25 & \textbf{25} & 0.0 & & & & \\
\texttt{le450\_25d} & 450 & 17425 & 25 & 27 & 31 & 27 & 29 & & & & \\
\texttt{le450\_5b} & 450 & 5734 & 5 & 5 & 12 & 7 & 7.2 & & & & \\
\texttt{le450\_5d} & 450 & 9757 & 5 & 8 & 18 & 8 & 15 & & & & \\
%\texttt{mug100\_1} & 100 & 166 & 4 & 4 & 4 & \textbf{4} & 0.0 & & & & \\
\texttt{mug100\_25} & 100 & 166 & 4 & 4 & 4 & \textbf{4} & 0.0 & & & & \\
%\texttt{mug88\_1} & 88 & 146 & 4 & 4 & 4 & \textbf{4} & 0.0 & & & & \\
\texttt{mug88\_25} & 88 & 146 & 4 & 4 & 4 & \textbf{4} & 0.0 & & & & \\
\texttt{mulsol.i.1} & 197 & 3925 & 49 & 49 & 63 & 50 & 0.0 & & & & \\
\texttt{mulsol.i.2} & 188 & 3885 & 31 & 48 & 58 & 48 & 0.1 & & & & \\
\texttt{myciel6} & 95 & 755 & 7 & 7 & 11 & \textbf{7} & 0.0 & & & & \\
\texttt{myciel7} & 191 & 2360 & 8 & 8 & 12 & \textbf{8} & 0.1 & & & & \\
\texttt{qg.order40} & 1600 & 62400 & 40 & 40 & 64 & 42 & 22 & \textbf{40} & 47 & & \\
\texttt{qg.order60} & 3600 & 212400 & 60 & 60 & 64 & 64 & 0.0 & \textbf{60} & 267 & & \\
%\texttt{queen11\_11} & 121 & 1980 & 11 & & 20 & & & 13 & 0.06 & & \\
%\texttt{queen12\_12} & 144 & 2596 & 12 & & 21 & & & 14 & 0.15 & & \\
%\texttt{queen13\_13} & 169 & 3328 & 13 & & 24 & & & 15 & 1.37 & & \\
%\texttt{queen14\_14} & 196 & 4186 & 14 & & 24 & & & 16 & 1.56 & & \\
%\texttt{queen15\_15} & 225 & 5180 & 15 & & 27 & & & 17 & 1.96 & & \\
%\texttt{queen16\_16} & 256 & 6320 & 16 & & 29 & & & 18 & 16.94 & & \\
\texttt{queen8\_12} & 96 & 1368 & 12 & 12 & 20 & \textbf{12} & 0.1 & & & & \\
\texttt{queen9\_9} & 81 & 1056 & 10 & 10 & 15 & \textbf{10} & 9.2 & & & & \\
\texttt{queen10\_10} & 100 & 1470 & 10 & 11 & 18 & 12 & 0.1 & 11 & 143 & & \\
\texttt{school1} & 385 & 19095 & 15 & 15 & 49 & \textbf{15} & 12 & & & & \\
\texttt{school1\_nsh} & 352 & 14612 & 14 & 14 & 40 & \textbf{14} & 14 & & & & \\
\texttt{wap01a} & 2368 & 110871 & 41 & 46 & 48 & 46 & 15 & & & & \\
\texttt{wap02a} & 2464 & 111742 & 40 & 44 & 49 & 47 & 18 & 44 & 83 & & \\
\texttt{wap03a} & 4730 & 286722 & 40 & 50 & 58 & 57 & 18 & 50 & 464 & & \\
\texttt{abb313GPIA} & 1557 & 53356 & 8 & 9 & 17 & 13 & 28 & 10 & 283 & & \\
\texttt{ash331GPIA} & 662 & 4181 & 3 & 4 & 8 & 4 & 2 & & & & \\
\texttt{ash608GPIA} & 1216 & 7844 & 3 & 4 & 10 & 4 & 12 & & & & \\
\texttt{ash958GPIA} & 1916 & 12506 & 3 & 4 & 10 & 5 & 11 & 4 & 41 & & \\
\texttt{will199GPIA} & 701 & 6772 & 7 & 7 & 9 & \textbf{7} & 2.2 & & & & \\
\multicolumn{12}{c}{}
\end{tabular}
\caption{Execution of \textsc{TabuEqCol} over benchmark instances}
\label{TABLE2}
\end{table}

\end{document}